\documentclass[aps,twocolumn,prl,showpacs,superscriptaddress,groupedaddress]{revtex4}  
\usepackage{dcolumn}
\usepackage{bm}
\usepackage{color}

\def\ltape{\hbox{\ $<$\hskip -8pt\raise -4pt\hbox{$\sim$}\ }}
\def\gtape{\hbox{\ $>$\hskip -8pt\raise -4pt\hbox{$\sim$}\ }}

   \ifx\pdfoutput\undefined
   \usepackage[dvips]{graphicx}
   \else
   \usepackage[pdftex]{graphicx}
   \fi
   \usepackage{epstopdf}
   \DeclareGraphicsExtensions{.pdf,.gif,.jpg,.eps,.png}
  
   \usepackage{amsmath}
  \usepackage{amssymb}

\begin{document}

\title{On the collisionless asymmetric magnetic reconnection rate}

\author{Yi-Hsin~Liu}
\affiliation{Dartmouth College, Hanover, NH 03750}
\author{M.~Hesse}
\affiliation{University of Bergen, Bergen, Norway}
\affiliation{Southwest Research Institute, San Antonio, TX 78238}
\author{P. A.~Cassak}
\affiliation{West Virginia University, Morgantown, WV 26506}
\author{M. A.~Shay}
\affiliation{University of Delaware, Newark, DE 19716}
\author{S.~Wang}
\affiliation{University of Maryland, College Park, MD 20742}
\author{L. -J.~Chen}
\affiliation{NASA-Goddard Space Flight Center, Greenbelt, MD 20771}
\affiliation{University of Maryland, College Park, MD 20742}

\date{\today}
\begin{abstract}

A prediction of the steady-state reconnection electric field in asymmetric reconnection is obtained by maximizing the reconnection rate as a function of the opening angle made by the upstream magnetic field on the weak magnetic field (magnetosheath) side. The prediction is within a factor of two of the widely examined asymmetric reconnection model [Cassak and Shay, Phys. Plasmas {\bf 14}, 102114, 2007] in the collisionless limit, and they scale the same over a wide parameter regime.  The previous model had the effective aspect ratio of the diffusion region as a free parameter, which simulations and observations suggest is on the order of 0.1, but the present model has no free parameters. In conjunction with the symmetric case [Liu et al., Phys. Rev. Lett. {\bf 118}, 085101, 2017], this work further suggests that this nearly universal number $0.1$, essentially the normalized fast reconnection rate, is a geometrical factor arising from maximizing the reconnection rate within magnetohydrodynamic (MHD)-scale constraints. 

\end{abstract}


\maketitle

{\it Introduction--} 
Magnetic reconnection at Earth's magnetopause not only allows the transport of solar wind plasmas into Earth's magnetosphere but also enhances the convection of magnetic flux to Earth's night side \cite{dungey61a}. The magnetic fields and plasma conditions on the two sides of the magnetopause current sheet are typically different [e.g., \citep{phan96a}]; a feature that also applies to current sheets in planetary \cite{masters15a,fuselier14b}, solar \cite{murphy12a}, laboratory \cite{yoo14a}, fusion \cite{mirnov06a} and turbulent \cite{servidio09a, zhdankin13a} plasmas. Reconnection with these different upstream conditions is commonly called {\it asymmetric}. To model the global circulation of magnetospheric plasmas around Earth and the magnetic energy release therein, it is crucial to understand how fast the magnetic flux is processed by asymmetric reconnection at Earth's magnetopause [e.g., \citep{borovsky08a,borovsky13a}]. 

One measure of the reconnection rate is the strength of the reconnection electric field inside the reconnection diffusion region which, according to Faraday's law, is proportional to the magnetic flux change rate at the diffusion region.
At Earth's magnetopause, directly measuring the reconnection electric field has been conducted although it remains challenging [e.g., \cite{LJChen17a, mozer07a, vaivads04a}]. A good proxy of the reconnection rate is the convective electric field upstream of the diffusion region induced by the inflowing plasma. 
Such an electric field was inferred from the ion velocity into the ion diffusion region [e.g., \cite{SWang15a, mozer10a, fuselier05a, phan01a}], or from the electron velocity into the electron diffusion region \cite{LJChen17a}. The reconnection rate can also be estimated from the magnitude of reconnected magnetic fields downstream of the ion diffusion region [e.g., \cite{phan01a}] using Sweet-Parker scaling \cite{sweet58a,parker57a}, or from the energy conversion rate \cite{rosenqvist08a}. 


Observational evidence \cite{SWang15a,fuselier16a, mozer10a} suggests that the strength of the reconnection electric field follows the scaling 
\begin{equation}
E_{CS}=2\left(\frac{B_1B_2}{B_1+B_2}\right)\left(\frac{V_{out}}{c}\right)\left(\frac{\delta}{L}\right)_{eff},
\label{CS}
\end{equation}
that is derived using conservation laws \cite{cassak07b}.
$B_1$ and $B_2$ are the reconnecting component of magnetic fields at the magnetosheath and magnetosphere sides, respectively. The outflow speed $V_{out}=(B_1B_2/4\pi {\bar \rho})^{1/2}$ is the hybrid Alfv\'en speed based on a hybrid density ${\bar \rho}=(B_1\rho_2+B_2\rho_1)/(B_1+B_2)$.  Here $(\delta/L)_{eff}$ is the {\it effective aspect ratio} of the diffusion region, which is a free parameter in this model for collisionless reconnection. Observations suggest that  $(\delta/L)_{eff}$ is of order 0.1.
Numerical simulations have also confirmed this scaling and demonstrated that $(\delta/L)_{eff}\sim 0.1$; these include local MHD simulations with a localized resistivity \cite{birn08a}, local two-fluid \citep{cassak08a} and local particle-in-cell (PIC) \citep{malakit10a,pritchett08a} simulations, as well as global magnetospheric MHD simulations \citep{borovsky08a, borovsky13a, quellette13a,BZhang16a,komar16a} and global Vlasov simulations \citep{hoilijoki17a}. This scaling along with $(\delta/L)_{eff}\sim 0.1$ was then employed to develop a quantitative model of the coupling between the solar wind and Earth's magnetosphere \cite{borovsky08a,borovsky13a}. Given these successes of Eq.~(\ref{CS}), it remains not understood why the {\it effective aspect ratio} in this model should be of order 0.1. This obviously requires an explanation. 

 

In this Letter, we provide a theoretical explanation for the collisionless asymmetric reconnection rate. We generalize the approach discussed in Ref.~\citep{yhliu17a} which was used to model the symmetric reconnection electric field. Through analyzing force-balance at the inflow and outflow regions, we cast the reconnection electric field into the form of a function of the opening angle made by the upstream magnetic field on the weak field side. A prediction is then obtained by maximizing this rate as a function of this opening angle, which we find to agree with $E_{CS}$ within a factor of two, with agreement in the scaling sense over a wide range of upstream plasma parameters. This comparison demonstrates that this nearly universal {\it effective aspect ratio} of order 0.1 in the collisionless limit \citep{cassak08a} is also the result of geometrical constraints on the MHD-scale, independent of the dissipation mechanism. \\



\begin{figure}
\includegraphics[width=7cm]{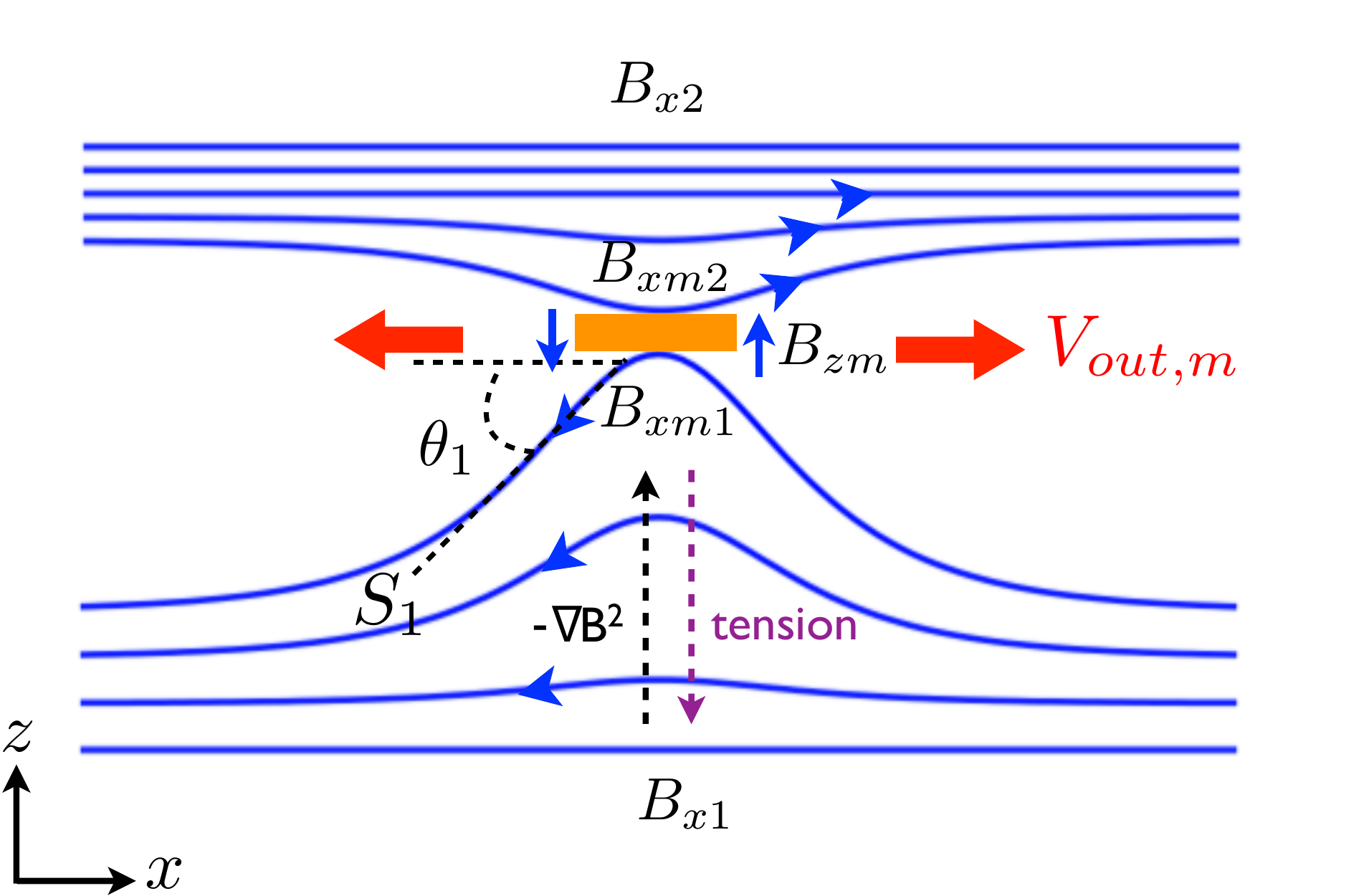} 
\caption {The geometry of magnetic fields upstream of the diffusion region for asymmetric reconnection. The orange box marks the diffusion region. $S_1=\mbox{tan}|\theta_1|$ marks the slope of the magnetic field line on side 1. The strength of the magnetic field is illustrated by the field line density.
}
\label{inflow}
\end{figure}

{ \it Constraint on the reconnecting field--} 
We consider the geometry and notation illustrated in Fig.~\ref{inflow}. The asymptotic field $B_{x2}$ on side 2 is larger than the asymptotic value $B_{x1}$ on side 1. Thus, sides 1 and 2 nominally correspond to typical conditions at the magnetosheath and magnetosphere, respectively. Unlike the model in Ref.~\cite{cassak07b}, the strength of the reconnecting field immediately upstream of the ion diffusion region can be different from the asymptotic field on each side. We use a subscript ``$m$'' in $B_{xmi}$ to indicate the {\it microscopic} ion-diffusion-region-scale, and $i=1,2$ indicates the two inflow sides. $V_{out,m}$ is the outflow speed immediately downstream of the diffusion region.
During the nonlinear stage of reconnection, the angle $\theta_i$ (as sketched for side 1) made by the upstream magnetic field lines opens out on each side. This geometry unavoidably induces a tension force ${\bf B}\cdot\nabla B_z/4\pi$ directed away from the x-line (as sketched for side 1), that is mostly balanced by the magnetic pressure gradient force $-(\nabla B^2/8\pi)_z$ directed toward the x-line (as sketched for side 1). Such a finite magnetic pressure gradient requires the reduction of the reconnecting magnetic field immediately upstream of the diffusion region. This effect is modeled in Ref.~\citep{yhliu17a} that results in an expression
\begin{equation}
B_{xmi}\simeq B_{xi}\frac{1-S^2_i}{1+S^2_i}.
\label{Bxm}
\end{equation}
Here $S_i=\mbox{tan}|\theta_i|$ is the slope of the upstream magnetic field line on each side, as sketched for side 1 in Fig.~\ref{inflow}.
From Eq.~(\ref{Bxm}), the reconnecting magnetic field $B_{xmi}$ vanishes as the opening angle approaches $45^\circ$ (i.e., $S_i \rightarrow 1$). 
In the 2D approximation, we can write  ${\bf B}=\nabla \times  A_y{\hat{\bf y}} + B_y \hat{\bf y}$. The sample field lines in Fig.~\ref{inflow} are evenly spaced contours of the flux function $A_y$, hence the ``line-density'' illustrates the strength of the in-plane magnetic field. The field lines approaching the diffusion region become less dense (i.e., weaker) compared with its asymptotic value on each side, illustrating the reduction of the reconnecting field due to the opening out of the upstream magnetic field lines.

The reconnected field immediately downstream of the diffusion region scales as
\begin{equation}
B_{zm}\simeq B_{xmi}S_i.
\end{equation}
This captures the trend that the opening angle made by the upstream magnetic field on side 1 is always larger, as illustrated in Fig.~\ref{inflow}. This also means that the reduction of reconnecting magnetic field on the weaker field side has a stronger effect in limiting the reconnection rate.

Because the field strength of the reconnecting field on side 1 is weaker than that on side 2, all possible solutions of this model must be found in the range $0< S_1 <1$. Therefore, we write $B_{zm}$ as a function of $S_1$:  
\begin{equation}
B_{zm}(S_1)\simeq \left(\frac{1-S_1^2}{1+S_1^2}\right)S_1 B_{x1}.
\label{Bz_S1}
\end{equation}

\begin{figure}
\includegraphics[width=8cm]{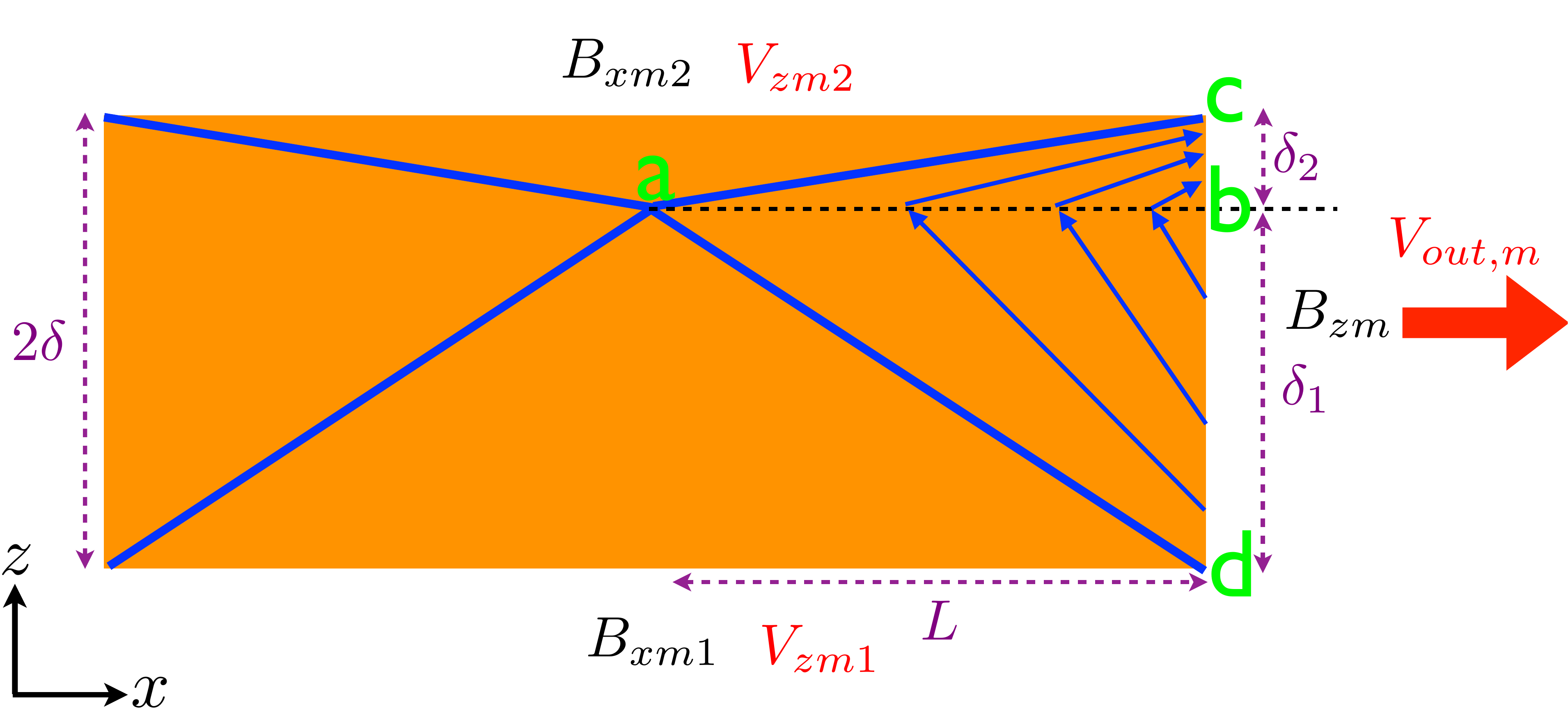} 
\caption {The geometry and dimension of the diffusion region. The strength of the magnetic field is illustrated by the field line density. Here $\delta_1+\delta_2=2\delta$. The label ``a'' marks the reconnection x-line.
}
\label{outflow}
\end{figure}

{ \it Constraint on the outflow speed--}
To estimate the reconnection electric field, $E_y\simeq B_{zm} V_{out,m}/c$, we need to calculate the outflow speed $V_{out,m}$. We consider the notation and geometry in Fig.~\ref{outflow}. The dimension of the diffusion region is $2L \times 2\delta$. Lines $a-c$ and $a-d$ represent the separatrices on side 2 and side 1 respectively, and ``$a$'' marks the x-line. We first derive the outflow density $\bar{\rho}$ as a result of mixing of plasmas from two sides. The integral form of Gauss' law for a 2D system is $\oint {\bf B}\cdot d{\bf l}=0$ where $d{\bf l}$ is along the perimeter of a closed 2D area. By applying this rule to the triangle area $a-b-c$ in Fig.~\ref{outflow} and  we get $\int_a^b B_z dx+\int_b^c B_x dz+\int_c^a{\bf B}\cdot d{\bf l}=0$. The last integral vanishes identically because the magnetic separatrix passes the upper right corner at point ``$c$''. Thus, $(B_{zm}/2) L\simeq (B_{xm2}/2)\delta_2$. A similar exercise reveals $B_{zm} L\simeq B_{xm1}\delta_1$. Combined with the relation $\delta_1+\delta_2=2\delta$, we get
\begin{equation}
B_{zm}=2\left(\frac{B_{xm1}B_{xm2}}{B_{xm1}+B_{xm2}}\right)\left(\frac{\delta}{L}\right).
\label{Bz}
\end{equation}
We now estimate the mass density as in Refs. \citep{cassak07b}, taking care to note that the conservation laws are evaluated at the microscopic ``$m$'' scale.  Mass conservation gives $2\bar{\rho}V_{out,m}\delta\simeq\rho_1 V_{zm1} L+\rho_2 V_{zm2}L$. In a 2D steady state, the out-of-plane electric field $E_y$ is uniform around the diffusion region and hence $V_{zm1}B_{xm1}=V_{zm2}B_{xm2}=V_{out,m}B_{zm}$. Eliminating the velocities gives the hybrid mass density \citep{cassak07b},
\begin{equation}
{\bar \rho}\simeq \frac{B_{xm1}\rho_2+B_{xm2}\rho_1}{B_{xm1}+B_{xm2}}.
\label{rho}
\end{equation}

Now we have enough information to derive the outflow speed from the momentum equation in the outflow direction $\hat{x}$, which is written as $(\rho/2)\partial_x V_x^2 \simeq B_z\partial_z B_x/4\pi -\partial_x B^2/8\pi$. Note that we have ignored the thermal pressure gradient by the same reasoning discussed in Ref.~\cite{birn10a}. To get an averaged outflow speed, we follow a process similar to Ref.~\citep{swisdak07a}; we apply $\int_0^Ldx\int_{-\delta}^{\delta} dz$ to the momentum equation, assuming $B_z=B_z(x)$, $B_x=B_x(z)$, $V_x=V_x(x)$ and an uniform density $\rho=\bar{\rho}$ inside the diffusion region. These lead to $(\bar{\rho}/2)V_{out}^22\delta \simeq (B_{zm}/2)L(B_{xm2}+B_{xm1})/4\pi-B_{zm}^22\delta/8\pi$. Substituting Eq.~(\ref{Bz}) for $B_{zm}$, we get
\begin{equation}
V_{out,m}\simeq \sqrt{\frac{B_{xm1}B_{xm2}}{4\pi \bar{\rho}}\left[1-4\frac{B_{xm1}B_{xm2}}{(B_{xm1}+B_{xm2})^2}\left(\frac{\delta}{L}\right)^2\right]}.
\label{Vout_dL}
\end{equation}
The first term inside the square brackets results from the averaged magnetic tension force and is the speed obtained in previous studies \citep{swisdak07a,cassak07b}. The reduction of the reconnecting field discussed in the previous section decreases the tension force that drives the outflow away from the diffusion region.
The second term proportional to $(\delta/L)^2$ is a new term that arises from the magnetic pressure gradient and it further reduces the outflow speed. However, the pre-factor dependent on $B_{xm1}$ and $B_{xm2}$ is 1 for the symmetric case \citep{yhliu17a, cassak17a} and decreases for increasing field asymmetries, so the correction to the outflow speed is weakened even more for asymmetric reconnection than symmetric reconnection \cite{yhliu17a}.  

We cast the outflow speed into a function of $S_1$($\simeq \delta_1/L$) instead,
\begin{equation}
{V_{out,m}}(S_1)\simeq \sqrt{\frac{B_{xm1}B_{xm2}-S_1^2B_{xm1}^2}{4\pi\bar \rho}}.
\label{Vout}
\end{equation}
The associated reconnection electric field is
\begin{equation}
E_y(S_1)\simeq B_{zm}(S_1)V_{out,m}(S_1)/c,
\label{rate}
\end{equation}
which is a function of $S_1$
using Eqs. (\ref{Bxm}), (\ref{Bz_S1}), (\ref{rho}) and (\ref{Vout}). We hypothesize that the reconnection rate corresponds to the maximum allowable value. Our prediction of the reconnection electric field is $E_R \equiv \mbox{max}(E_y)$, that can be found in the range $0 \leq S_1 \leq 1$. 

Note that writing $B_{xm2}$ as an explicit function of $S_1$ can be done, but there is no simple expression for it. We need to use the relation $B_{xm2}S_2=B_{xm1}S_1$ and Eq.~(\ref{Bxm}) to derive $S_2(S_1)$ first, which involves finding the roots of a cubic function $S^3_2+ [B_{zm}(S_1)/B_{x2}] S^2_2-S_2+B_{zm}(S_1)/B_{x2}=0$. $S_2(S_1)$ is then plugged into Eq.~(\ref{Bxm}) to get $B_{xm2}(S_1)$. These calculations can be performed numerically in a straightforward fashion. 




 \begin{figure}
\includegraphics[width=9cm]{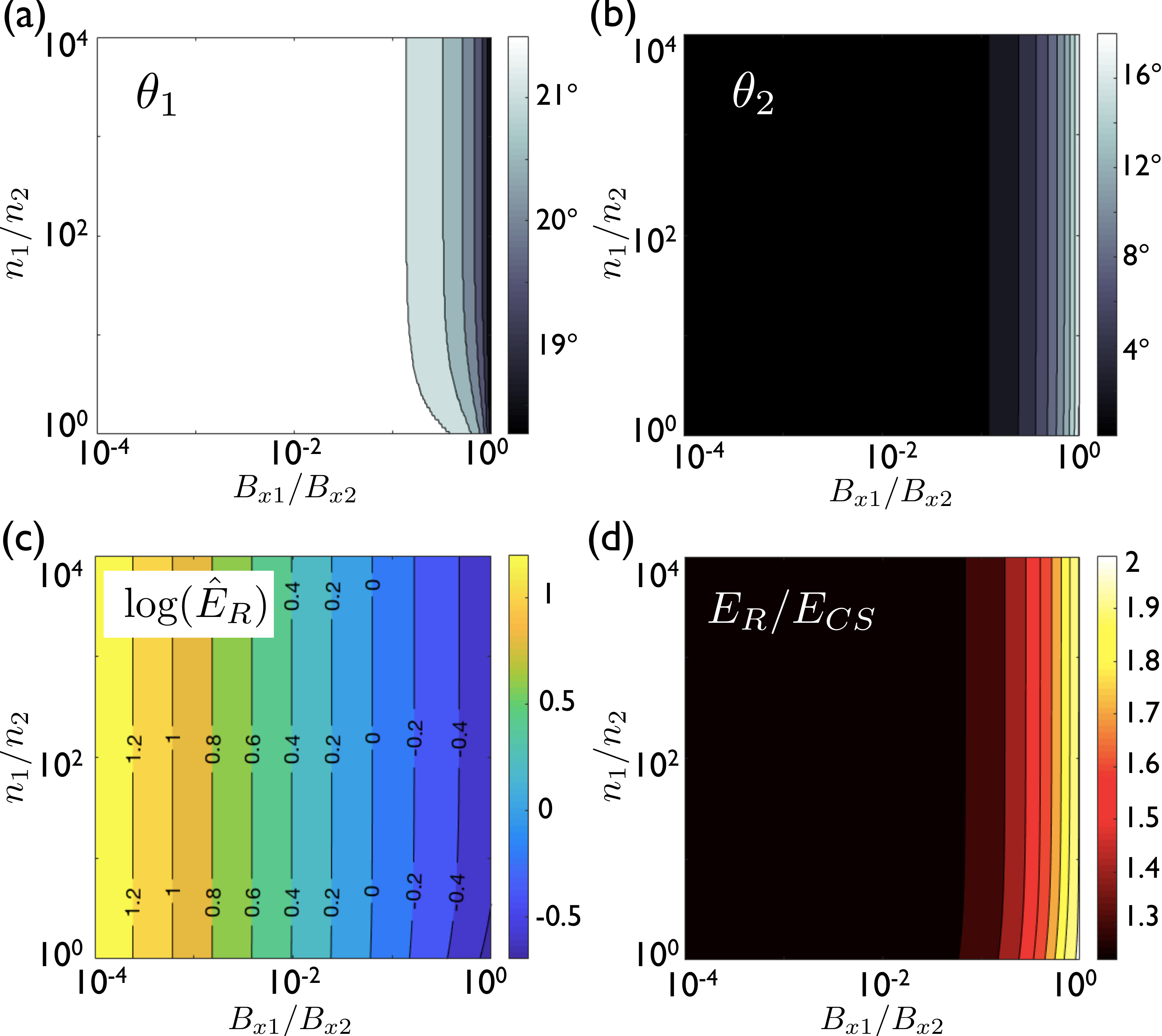} 
\caption {(a) The predicted opening angle made by the upstream magnetic field on side 1 is plotted as a function of $B_{x1}/B_{x2}$ and $n_1/n_2$. (b) The predicted opening angle made by the upstream magnetic field on side 2. (c) The contour of the predicted reconnection electric field normalized by the side 1 (magnetosheath) value. (d) The ratio of the predicted reconnection electric field to the prediction in Eq.~(\ref{CS}) assuming $\delta / L = 0.1$. 
}
\label{scaling_2D}
\end{figure}

{ \it Prediction--}
In the following, we find the maximum reconnection electric field $E_R$ from Eq.~(\ref{rate}) numerically. The result for a wide parameter range of magnetic field ratio $B_{x1}/B_{x2}$ and density ratio $n_1/n_2$ is shown in Fig.~\ref{scaling_2D}. The predicted opening angles on the two sides of the current sheet are shown in Fig.~\ref{scaling_2D}(a) and (b). The opening angle $\theta_1$ of the upstream magnetic field line on side 1 increases mildly from from $\simeq 18.2^\circ$ in the symmetric limit to  $\simeq 21.5^\circ$ in the strong field asymmetry limit. In the same limit, the field line opening angle $\theta_2$ on side 2 becomes small ($\rightarrow 0^\circ$) because the magnetic field is much stiffer on side 2 compared to that on side 1. This qualitatively agrees with all previous asymmetric reconnection simulations, which show $\theta_1 > \theta_2$. In Fig.~\ref{scaling_2D}(c), the reconnection electric field $\hat {E}_R\equiv c E_R/V_{Ax1}B_{x1}$ is normalized to the Alfv\'en speed $V_{Ax1}\equiv B_{x1}/\sqrt{4\pi \rho_1}$ and the field strength $B_{x1}$ at the magnetosheath (side 1). The normalized rate $\hat{E}_R$ is $\simeq 0.2$ in the symmetric limit (i.e., $\mbox{log}(\hat{E}_R)\simeq -0.7$ when $n_1/n_2=1$ and $B_{x1}/B_{x2}=1$), as expected from Ref.~\citep{yhliu17a}. In Fig.~\ref{scaling_2D}(d), we compare our prediction to $E_{CS}$ with $(\delta/L)_{eff}=0.1$. It is important to learn that this prediction agrees with $E_{CS}$ within a factor of two and they scale together over a wide range of parameter space. In conjunction with the symmetric case discussed in Ref. \cite{yhliu17a}, this consistency in the asymmetric limit suggests that the geometrical factor, $(\delta/L)_{eff}\simeq 0.1$, left unexplained in Eq.~(\ref{CS}), also arises from the MHD-scale constraints imposed at the inflow and outflow region. 

\begin{figure}
\includegraphics[width=9cm]{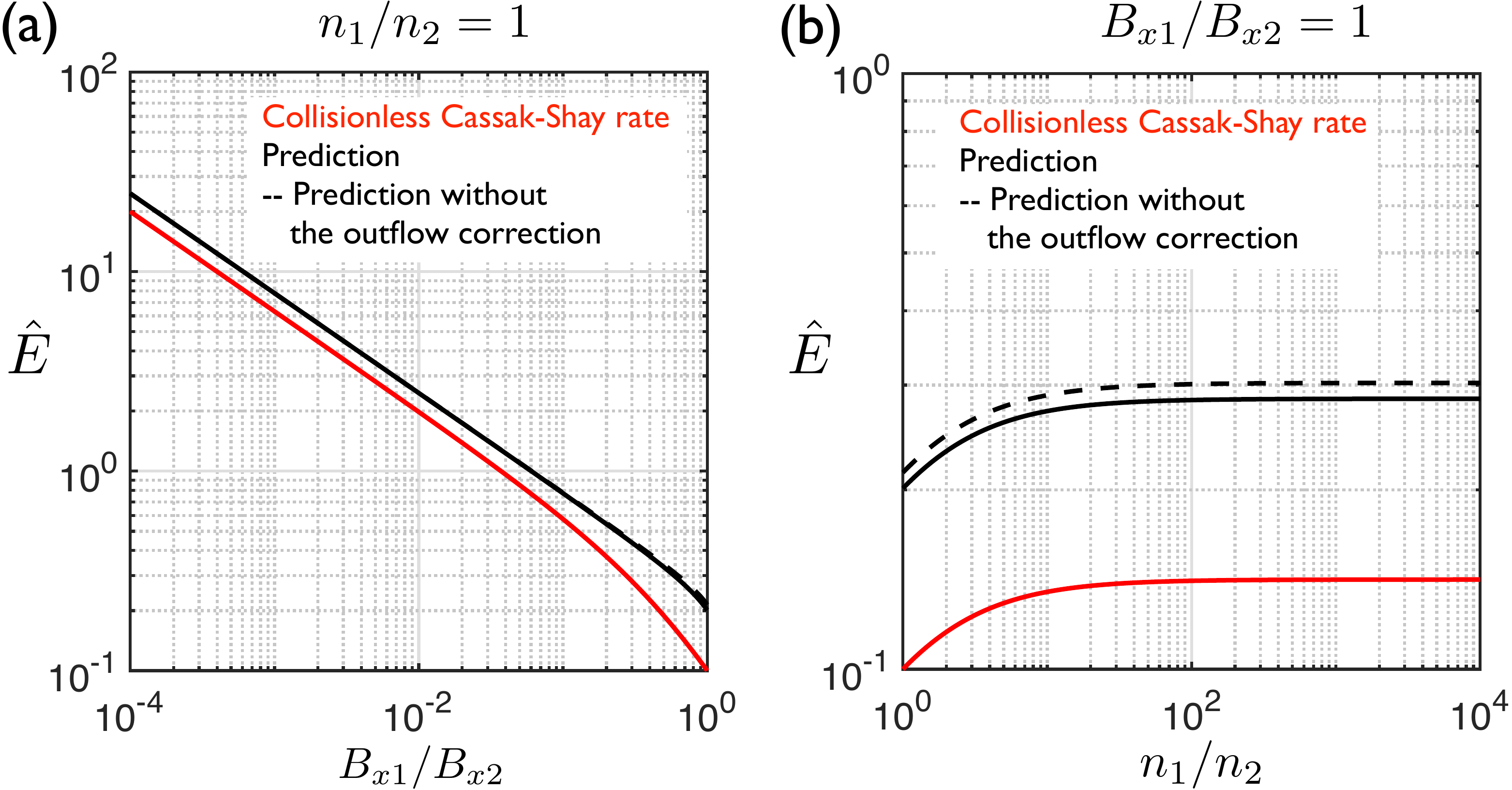} 
\caption {The predicted reconnection electric field normalized by the side 1 (magnetosheath) conditions is plotted as a function of $B_{x1}/B_{x2}$ with a symmetric density in (a) and as a function of $n_1/n_2$ with a symmetric reconnecting magnetic field in (b). The prediction is shown in solid back, the prediction of Eq.~(\ref{CS}) in red, and the prediction without the outflow correction in dashed black. 
}
\label{scaling}
\end{figure}

To understand better the difference in different models, we plot the predictions as a function of $B_{x1}/B_{x2}$ with a fixed $n_1/n_2=1$ in Fig.~\ref{scaling}(a) and as a function of $n_1/n_2$ with a fixed $B_{x1}/B_{x2}=1$ in Fig.~\ref{scaling}(b). Red curves show the value of $E_{CS}$ normalized to $V_{Ax1}B_{x1}/c$, solid black curves are $\hat{E}_R$ (our prediction), dashed black curves are the maximum of Eq.~(\ref{rate}) using ${V_{out,m}}(S_1)\simeq (B_{xm1}B_{xm2}/4\pi\bar \rho)^{1/2}$ instead of Eq.~(\ref{Vout_dL}); i.e., the reduction of the outflow speed from the magnetic pressure gradient is not considered. The red and solid black curves exhibit a similar scaling, as suggested in Fig.~\ref{scaling_2D}(d). The dashed black curve is very close to the solid black curve in each panel, suggesting that  the reduction of the reconnecting field, rather than the reduction in $V_{out,m}$ due to the magnetic pressure gradient force, is the dominant mechanism that constrains the rate.

{ \it Summary and discussion--} 
In this Letter, we derive the collisionless asymmetric magnetic reconnection rate using a new approach. The prediction is obtained through maximizing a model rate that considers the MHD-scale constraints at both the inflow and outflow regions. The predicted value is found to be within factor of two of the collisionless asymmetric reconnection rate that was widely examined \citep{cassak08a,cassak07b}. This comparison suggests that constraints at the MHD-scale explain the geometrical factor $(\delta/L)_{eff}$ of order 0.1 inferred but not explained in the rate model of Ref.~\citep{cassak08a}, putting the scaling in Eq.~(\ref{CS}) on solid footing. The analysis further shows that the dominant limiting effect that constrains the maximum reconnection rate is the field reduction at the weak field (magnetosheath) side.

However, caveats need to be kept in mind when applying this theory. 
An out-of-plane guide field does not affect the in-plane tension force but can contribute to the magnetic pressure gradient in the force balance. The same prediction applies to a general case with a guide field only if the reconnection process does not significantly alter the guide field strength near the x-line. The normalized rate remains to be $\sim 0.1$ in the strong guide field limit, at least, for symmetric cases \citep{yhliu14a}.
This model does not include the effect of the diamagnetic drift driven by the combination of the pressure gradient across the sheet and a finite guide field. The diamagnetic drift can suppress magnetic reconnection \citep{yhliu16a,beidler11a,swisdak10a,swisdak03a}. In addition, flow shear commonly present at the flank of the magnetopause can also reduce the reconnection rate \citep{doss15a,cassak11a}. Potential 3D and turbulence effects \citep{ergun16a, price16a, le17a, yhliu15b, daughton14a} are not included in this 2D analysis. Finally, while this theory works in most models, including PIC, hybrid, two-fluid models and MHD with a localized resistivity, it does not apply to MHD systems with a uniform resistivity; i.e., a uniform resistivity does not seem to support the maximum rate allowed by the constraints imposed at the upstream and downstream regions.

Nevertheless, by comparing with the well-established scaling \cite{cassak08a,cassak07b} previously found in the asymmetric limit of collisionless plasmas, the consistency demonstrated in this Letter confirms the capability of this new approach \cite{yhliu17a,cassak17a} in explaining the fast reconnection rate in a more general configuration. This result is timely to the study of collisionless magnetic reconnection in Earth's magnetosphere. The high cadence electric field measurement on board of NASA's Magnetospheric Multiscale spacecrafts (MMS) and their close deployment provide an invaluable opportunity to study the reconnection rate \cite{LJChen17a} and perhaps the effective aspect ratio of diffusion region in both the magnetopause and magnetotail. \\

\acknowledgments Y.-H. Liu thanks M. Swisdak and J. Dahlin for
helpful discussions. Y.-H. L. is supported by NASA grant NNX16AG75G and MMS mission. M. H. is supported by the Research Council of Norway/CoE under contract 223252/F50, and by NASA's MMS mission. P. A. C. is supported by NASA grant NNX16AF75G and NSF grant AGS1602769. M. S. is supported by NASA grants  NNX08A083G-MMS IDS, NNX17AI25G. S. W. and L.-J. C. are supported by DOE grant DESC0016278, NSF grants AGS-1202537, AGS-1543598 and AGS-1552142, and by the NASA's MMS mission.\\ 


\end{document}